\documentclass[aip,amsmath,amssymb]{revtex4-2}
\usepackage[dvipdfmx]{graphicx}
\usepackage[hang,small,bf]{caption}
\usepackage[subrefformat=parens]{subcaption}
\everymath{\displaystyle}
\begin{document}
\title{Effect of surfactants on the elasticity of the liquid--liquid interface}

\author{Shunta Kikuchi and Hiroshi Watanabe}
\affiliation{
    Department of Applied Physics and Physico-Informatics, Keio University, Yokohama, Kanagawa 233-8522, Japan
}

\email{hwatanabe@appi.keio.ac.jp}

\begin{abstract}
    We investigated the effect of surfactants on an interface between two kinds of liquid by molecular dynamics simulation. We adopted the simple bead-spring model with two atoms as the surfactants. We controlled the interfacial tension of the surfactant adsorbed on the interface by changing the bond length. Although the interface's structure changed depending on the magnitude of the interfacial tension, the interface was stable even under conditions where the interfacial tension was virtually zero. The Fourier spectrum of the fluctuations of the surface structure showed a crossover from $q^2$ to $q^4$ when the interfacial tension was almost zero, where $q$ is the wavenumber. This crossover means that the bending rigidity is dominant for the restoring force when the surfactant molecules are sufficiently absorbed on the interface and the interfacial tension is almost zero, whereas the interfacial tension is dominant when the interfacial tension is a finite value.
\end{abstract}

\maketitle

\section{Introduction}

Surfactants are amphiphilic particles with hydrophilic and hydrophobic groups and have various uses. For example, they are used in detergents and gas hydrate accelerators \cite{zhong2000surfactant}. In these applications, surfactants accumulating at interfaces decrease the interfacial tension. Therefore, it is very important in engineering applications to investigate the effects of surfactants on interfaces. Through experiments and simulations, many studies have been conducted on the effects of various surfactants on interfaces. In experiments, the pendant drop method is mainly used to measure the interfacial tension. Recently, the interfacial tension can be measured with high accuracy by using a high-speed camera\cite{tang2020effects}. However, it is challenging to observe microscopic phenomena through experiments, such as nanobubble formation or the behavior of molecules near an interface. With molecular dynamics (MD) simulations, the interfacial tension is determined using a microscopic stress tensor. MD simulations enable us to simulate systems that are difficult to observe experimentally and to determine which parts of molecules are essential\cite{fan2020molecular,derreumaux2007coarse}. Various studies of surfactants have been conducted by MD simulations, such as those on the adsorption of gas hydrates and the dependence of the hydrophobic group lengths\cite{choudhary2018effect,smit1990effects}.

Recently, the properties of lipid bilayers and membranes have been investigated extensively\cite{goetz1999mobility,Illya2005,Shinoda2010,watson2012determining,shiba2016monte}. A critical topic of the membranes is the bending rigidity and fluctuations in lipid bilayers. According to the continuum theory by Helfrich, the fluctuation in lipid bilayers contains two parts, the interfacial tension and bending rigidity~\cite{helfrich1973elastic}. The interfacial tension involves $q^2$, and the bending rigidity involves $q^4$ fluctuation with the wave number $q$. It was reported that the $q^4$ term is dominant for the dimyristoylphosphatidylcholine lipid bilayers~\cite{Brandt2011}. However, it is unclear under what conditions $q^4$ becomes dominant. One important parameter in surfactants is the length of the hydrophobic group. Several studies reported changes in the behavior of surfactants depending on the length of the hydrophobic group~\cite{Shi2015, Wu2016}. However, the longer hydrophobic group yields entropy effects, complicating the phenomenon and making analysis difficult. Therefore, we adopt a simple model to investigate the fluctuation of the membrane formed by surfactants and clarify its mechanical properties.

This study aims to determine when $q^4$ fluctuation appears and dominates in the membrane formed by surfactants. For simplicity, we study the monolayer of the surfactants absorbed at the liquid-liquid interface rather than the bilayer. We adopted the simple bead-spring model with two atoms as the surfactants. As shown later, we found that the interfacial tension of the liquid-liquid interface strongly depends on the bond length of the surfactants. There is a region where the interfacial tension is almost zero at a certain concentration of a surfactant and bond length. By utilizing this fact, we controlled the interfacial tension of the surfactant adsorbed on the interface by changing the bond length. The interface structure was maintained, but the properties fluctuated when the interfacial tension was virtually zero. On the other hand, when the interfacial tension was finite, the interface was flat, and the surfactant molecules were seeping into the bulk. The crossover from $q^2$ to $q^4$ fluctuation was observed when the interface tension vanished. This crossover indicates that when the interfacial tension is finite, the interfacial restoring force originates from the interfacial tension. In contrast, it originates from the bending rigidity when the interfacial tension is almost zero.

The rest of the paper is organized as follows. In section II, we describe the method. The results are described in Sec. III. Section IV is devoted to the summary and discussion.

\section{Method}

We adopt the LJ potential and Weeks--Chandler--Andersen (WCA) potentials to model a binary liquid and a surfactant molecule\cite{lennard1931cohesion,weeks1971role}. The potential function $\phi$ is given by
\begin{equation}
    \phi(r)=\left\{
    \begin{array}{ll}
        4 \varepsilon \left \{ \left  ( \frac{\sigma}{r} \right )^{12} - \left ( \frac{\sigma}{r} \right )^{6} \right \} & (r < r_c),     \\
        0                                                                                                                & (r \geq r_c ),
    \end{array}
    \right.
    \label{eq:lj}
\end{equation}
where $r$ is the distance between atoms, $r_c$ is the cut-off length, $\varepsilon$ is the well depth, and $\sigma$ is the diameter of atoms. Hereafter, all physical quantities are expressed in units of length $\sigma$, energy $\varepsilon$, and so forth. The cut-off length is set as $r_c=3.0$ for the LJ potential. We adopt $r_c = 2^{1/6}$ for the WCA potential so that the potential function exhibits only a repulsive force.

We consider the system in which the surfactant molecules are absorbed on the interface of a binary liquid. Surfactants act as amphiphilic molecules.
\begin{figure}[htbp]
    \includegraphics[width=15cm]{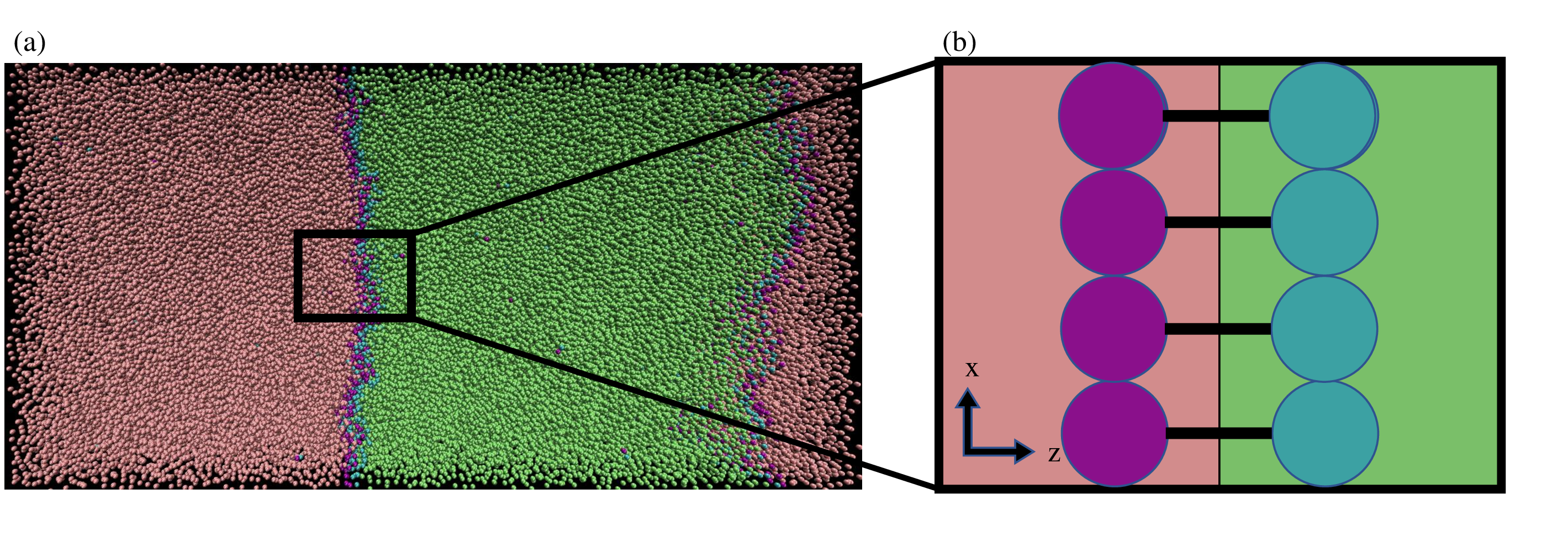}
    \caption{(Color online) (a) Snapshot of the system. For visibility, a small system with N = 200000  is shown. (b) Schematic view of the interface. Surfactant molecules are absorbed on the interface between two types of liquid. The interface is normal to the $z$-axis.}
    \label{fig:hosonaga_model}
\end{figure}
A binary liquid is modeled as a monatomic molecule. The two types of atom that make up a binary liquid are called A and B atoms, respectively. The interactions between atoms of the same types are LJ interactions and those between atoms of different types atoms are WCA interactions. The surfactant is modeled as a diatomic molecule, that is, it contains A and B atoms. We chose conditions under which the two liquids separate without mixing.
We adopt the interaction length $1.05 \sigma$ between the A and B atoms of surfactants to represent that the surfactants are slightly larger than the solvents. We adopt the harmonic potential $E(r)$ as the surfactant bond force as
\begin{equation}
    E(r)=K(r-l_0)^2,
    \label{eq:harmonic}
\end{equation}
where $K$ is the constant and $l_0$ is the bond length. We considered two cases, the weak bond $K=10$ and the strong bond $K=200$. WCA interactions between atoms of the same surfactant molecule are not considered. The bond length range is $0.50 \leq l_0 \leq 1.50$. The mass of all atoms is set to $m = 1.0$.

A typical snapshot of the simulation is shown in Fig.~\ref{fig:hosonaga_model}(a). We use LAMMPS (Large-scale Atomic/Molecular Massively Parallel Simulator) for the numerical integration of the equation of motion~\cite{LAMMPS1995,LAMMPS2022}. Visual molecular dynamics (VMD) was used for visualization\cite{humphrey1996vmd}. The simulation box is a cuboid of size $L_x\times L_y \times L_z$, where $L_x = 100, L_y = 100$, and $L_z = 200$. The concentration $\rho$ is defined as
\begin{equation}
    \rho = \frac{N}{L_xL_yL_z} = \frac{N}{V},
    \label{eq:concentration}
\end{equation}
where $N$ is the number of particles and $V$ is the volume of the system. We set $\rho=0.8$ and $N=1~600~000$. We perform the $NVT$ ensemble and control the temperature using a Langevin thermostat. The concentration $\rho_s$ of the surfactants is defined by the number concentration of the surfactants relative to the total volume. The temperature $T$ is fixed at $T = 1.0$ throughout simulations. The velocity Verlet algorithm is used for time integration up to $3~000~000$ with a time step of $0.005$. The surfactants are initially placed at the liquid-liquid interface and remain at the interface after sufficient relaxation as shown in Fig.~\ref{fig:hosonaga_model}~(b).

After the system was sufficiently thermalized, the interfacial tension $\gamma$ was measured. Since the interface is normal to the $z$ axis, the interfacial tension $\gamma$ is given by
\begin{equation}
    2L_{x}L_{y}\gamma = \int (P_{zz}-\frac{P_{xx}+P_{yy}}{2})dV = \int (P_{zz}-\frac{P_{xx}+P_{yy}}{2}) L_{x}L_{y}dz ,
    \label{eq:surface_tension}
\end{equation}
where $P_{xx}$, $P_{yy}$, and $P_{zz}$ are the diagonal components of the stress tensor\cite{allen2017computer}. Note that factor $2$ originates from the periodic boundary condition.

We measure the spectra of the fluctuations of the interface structure to investigate the origin of the restoring force at the interface. In this study, the interface was divided into grids of $30\times 30$ at $80<z<120$. We determined the center of the gravity of the surfactants in each grid and calculated the interface height $h(x,y)$. The form of spectrum $|h(q)|^2$ is expected to be\cite{goetz1999mobility}
\begin{equation}
    |h(q)|^2 = \frac{k_B T}{\gamma q^2+\kappa q^4},
    \label{eq:fluctuation}
\end{equation}
where $q$ is the wavenumber, $h(q)$ is the Fourier transform of $h(x,y)$, $\kappa$ is the bending rigidity and $k_B$ is the Boltzmann constant. The Boltzmann constant $k_b$ is set to be unity in our simulations. The statistical error was estimated using 15 independent samples. To observe the crossover from fluctuations dominated by the interfacial tension to those dominated by the bending rigidity, we set the surfactant concentration $\rho = 0.025$ at which the surfactant tension is virtually zero.

\section{Results}

\subsection{Bond length dependence of interfacial tension}

\begin{figure}[htbp]
    \begin{center}
        \includegraphics[width=10cm]{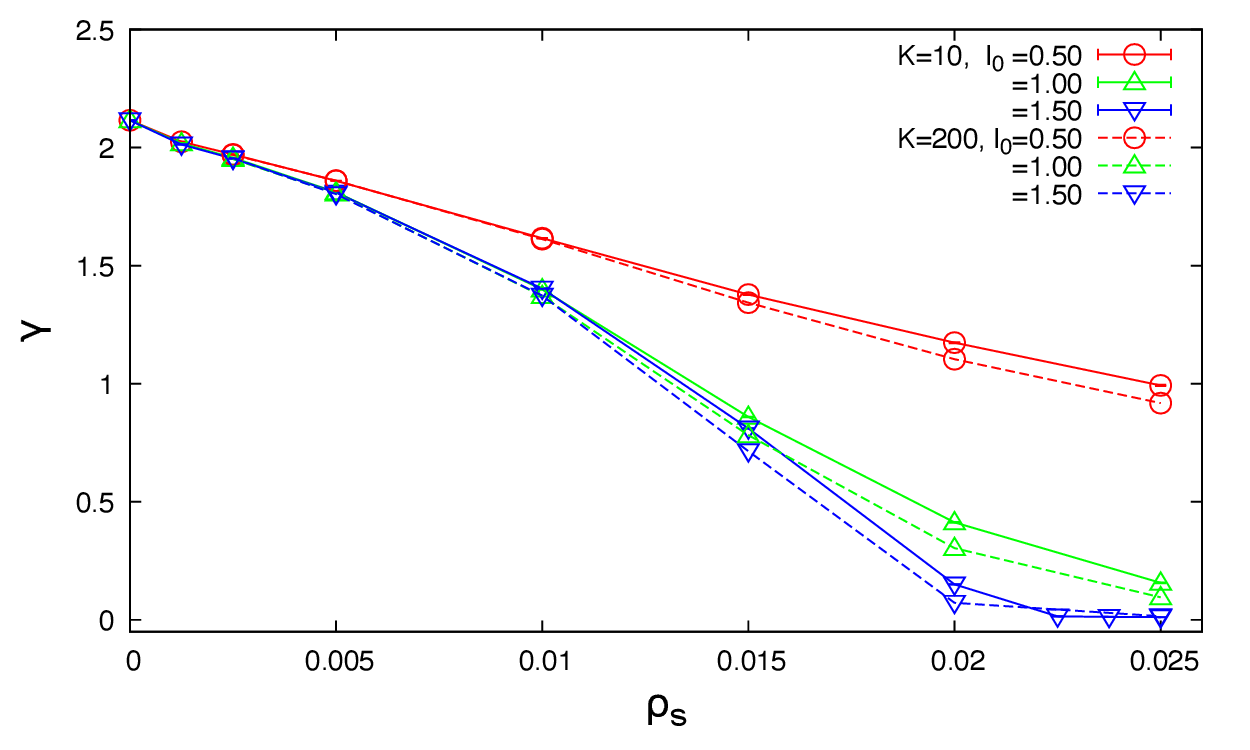}
        \caption{Concentration dependence of interfacial tension at $l_0 = 0.50, 1.00$, and $1.50$. The solid line represents the results for the weak spring constant $K=10$, and the dashed line represents the results for the strong spring constant $K=200$. Statistical errors are smaller than the symbol size. Concentration dependence shows that the bond length affects the interfacial tension.}
        \label{fig:length_itf}
    \end{center}
\end{figure}

We investigated how the bond length and the concentration of the modeled surfactant affect the interfacial tension. The dependence of the interfacial tension on the surfactant concentration is shown in Fig.~\ref{fig:length_itf}. The interfacial tension is affected by the bond length. The spring constant has little effect on the results and the bond length is the dominant parameter.Hereafter, we define the saturation concentration at which the interfacial tension becomes almost zero. Since the spring constant was found to have little effect on the results, we adopted the weak bond $K=10$ for all the following simulations for the computational convenience, since the strong bond requires the shorter time step.

The bond length dependence on the surface tension is shown in Fig.~\ref{fig:length_itf}. The surfactant at a low concentration ($\rho_s=0.015$) and a high concentration ($\rho_s=0.025$) are shown. When the concentration of the surfactants is low, the interfacial tension behaves in a convexly downward trend, and reaches the minimum when the bond length is around 1.2. When the concentration is high, the interfacial tension is finite for a small bond lengths but is virtually zero for a sufficiently large bond length.

\begin{figure}[htbp]
    \centering
    \includegraphics[width=10cm]{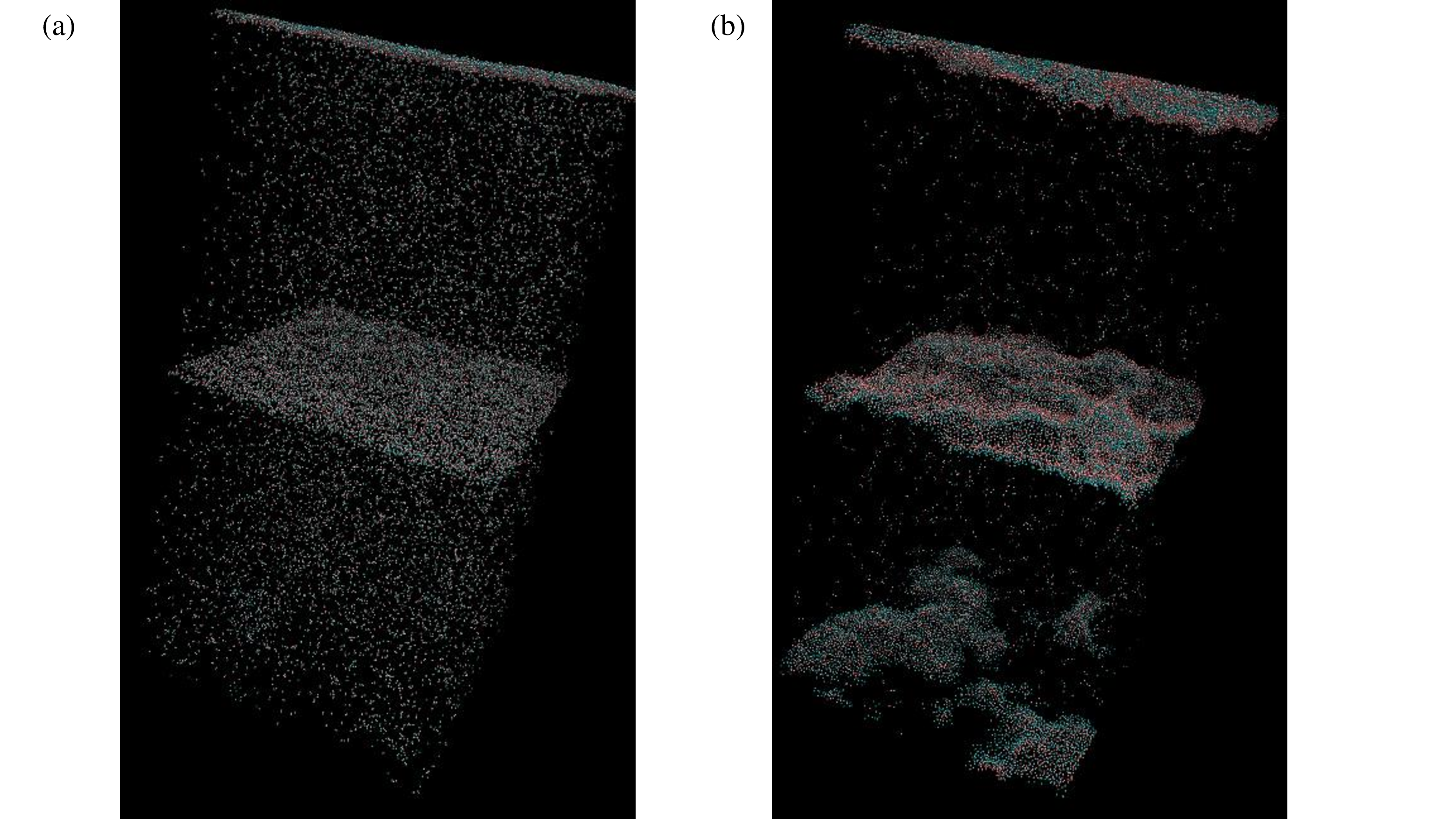}
    \caption{(Color online) Snapshots of interfacial structure for $\rho_s = 0.025$ and $T = 1.0$. (a) $l_0 = 0.50$. (b) $l_0 = 1.50$. Only surfactants are shown.}
    \label{fig:vmd}
\end{figure}

\begin{figure}[htbp]
    \begin{center}
        \includegraphics[width=10cm]{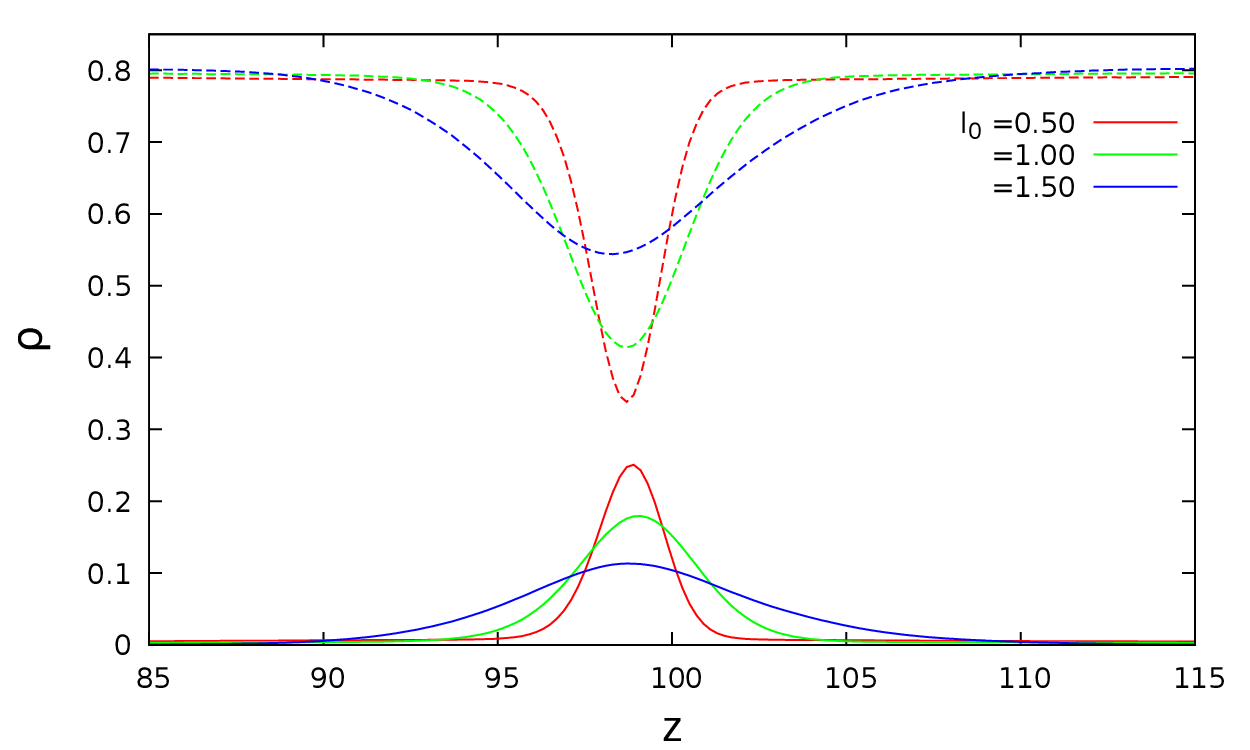}
        \caption{(Color onine) Surfactant concentration and solvent density profiles. The concentrations of the sufractant are shown with the solid lines and those of the solvent are shown with the dashed lines. As the bond length increases, the interface width also increases, but the surfactant remains adsorbed at the interface.}
        \label{fig:rho}
    \end{center}
\end{figure}

The snapshots of the interfacial structure are shown in Fig.~\ref{fig:vmd}. Figure~\ref{fig:vmd}(a) is the case where the system exhibits a finite interfacial tension ($l_0=0.50$). The interface is flat and does not fluctuate. In addition, the surfactant seeps into the solvent. In contrast, the interface highly fluctuates when the interfacial tension is almost zero (Fig.\ref{fig:vmd}(b)). Although the interface is highly fluctuating, the interfacial structure is maintained. The surfactant concentration and solvent density profiles are shown in Fig.~\ref{fig:rho}. Even when the interfacial tension is almost zero, the surfactant is localized near the interface, and the interfacial width remains finite.

\begin{figure}[htbp]
    \begin{center}
        \includegraphics[width=10cm]{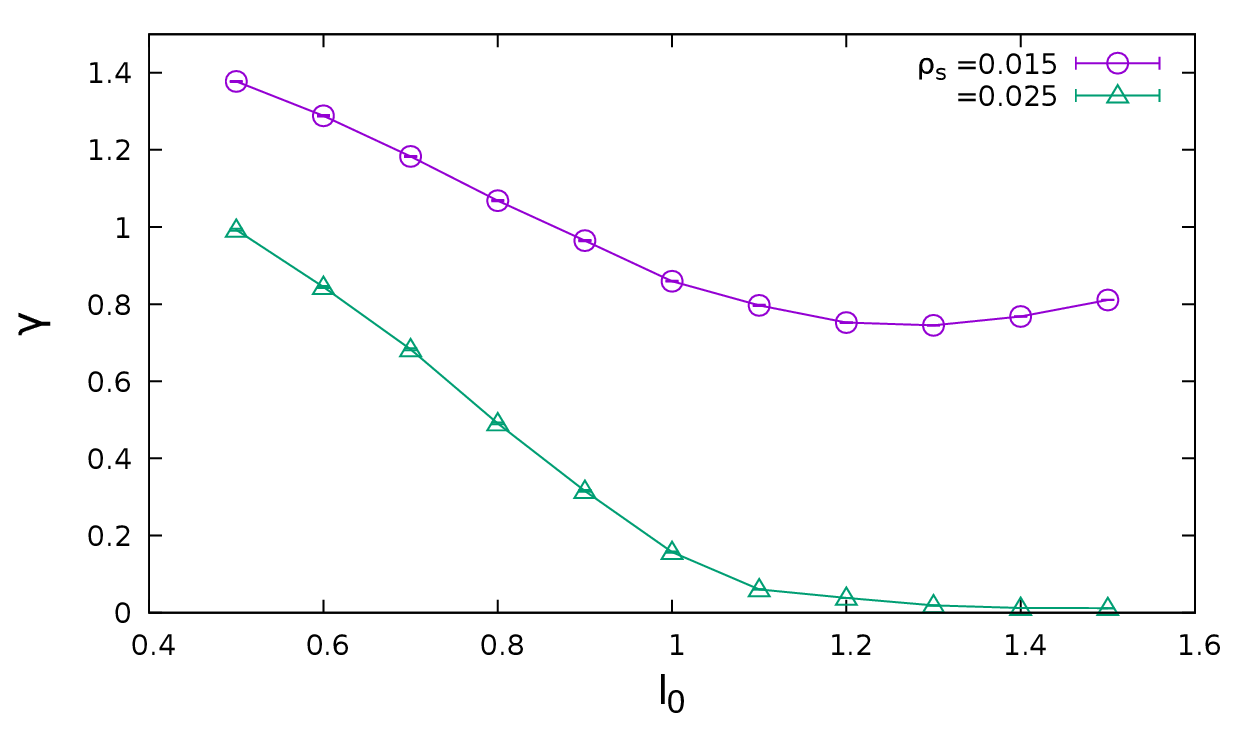}
        \caption{Bond length dependence of interfacial tension $\gamma$ for $\rho_s = 0.015$ and $0.025$. For $\rho_s = 0.015$, the interfacial tension is a minimum around $l_0 = 1.2$. For $\rho_s = 0.025$, the saturation concentration changes. The bond length between atoms of the surfactant molecule affects the interfacial tension.}
        \label{fig:length0.015_0.025}
    \end{center}
\end{figure}

\begin{figure}[htbp]
    \begin{center}
        \includegraphics[width=10cm]{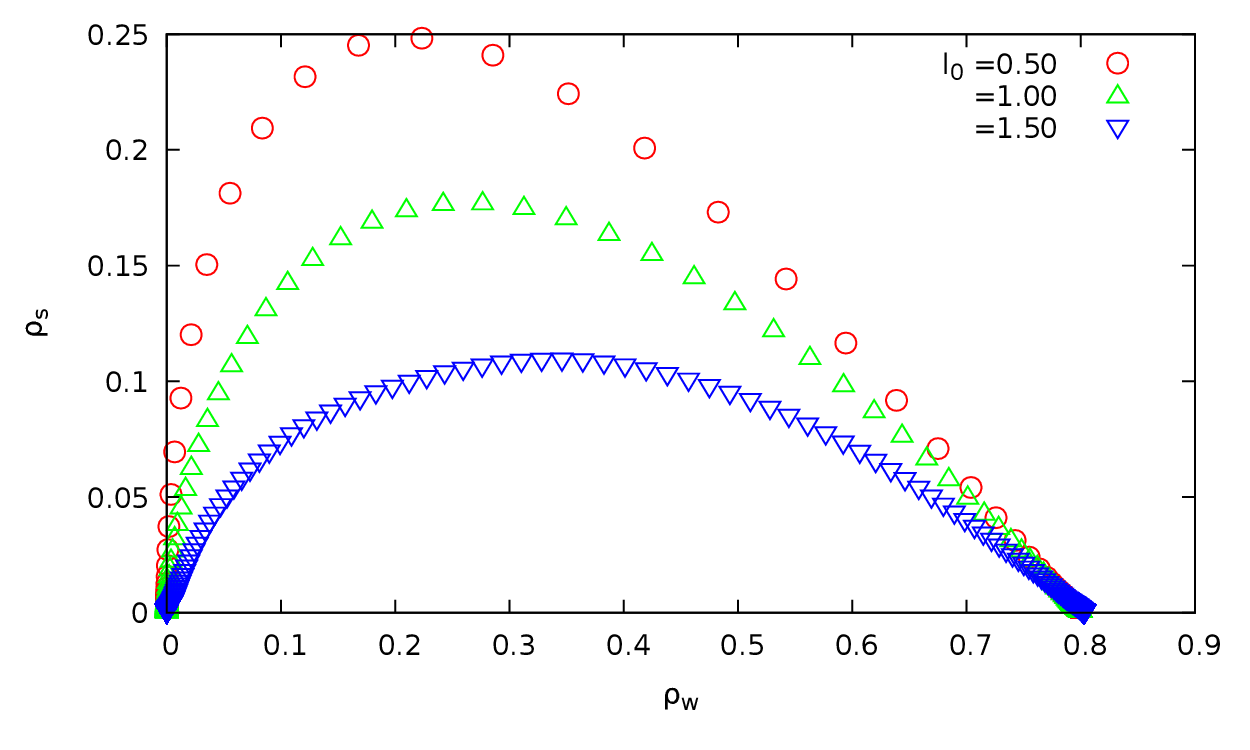}
        \caption{The solvent-density dependence on the surfactant concentration. While the average solvent density is $0.8$, the highest surfactant concentration is in the region of low solvent density, around $0.2$ to $0.35$ depdending on the bond length, indicating that the surfactants favor low solvent density.}
        \label{fig:sur_water_rho}
    \end{center}
\end{figure}

Figure~\ref{fig:rho} suggests that surfactants favor low solvent density. It is known that solutes adsorb at the interface even if the liquid is poor solvent because the interface between two liquids that repel each other has a lower density~\cite{Taddese2015}. The similar phenomenon can happen in the surfactants-solvent system. To illustrate this, we investigated the density dependence of the surfactant concentration on the density of the solvent. Since the system is uniform in the $x$ and $y$ directions, the local density is only $z$ dependent. Suppose the local density of A and B atoms are $\rho_A(z)$ and $\rho_B(z)$. Then the local density of the solvent is defined by $\rho_\mathrm{w}(z) =\rho_A(z) + \rho_B(z)$. We also determined the local density of the surfactants $\rho_\mathrm{s}(z)$, and plot the points $(\rho_\mathrm{w}(z), \rho_\mathrm{s}(z))$ for various position at $z$. The solvent-density dependence of the surfactant concentration is shown in Fig.~\ref{fig:sur_water_rho}. As shown in the figure, the surfactants favor low solvent density. Since the two liquid phases repel each other, the density of solvent near the interface is low. One can be seen that the surfactant concentration is higher where the solvent phase density is lower. Therefore, the lower interfacial tension increases the low-density region of the solvent at the interface, which meands more surfactant to be adsorbed at the interface. This is why the surfactant is not dissolved in the solvent in Fig.~\ref{fig:vmd}~(b). Wu \textit{et al.}~reported that the diffusion coefficient in lipid bilayers increases as the length of alkyl groups of phospholipid decreases~\cite{Wu2016}. A change involves this phenomenon in the attraction between lipids depending on the length of the alkyl groups. This is consistent with our results that the diffusion coefficient of surfactants increases when their bond length is short, and consequently, they seep into the bulk.

\begin{figure}[htbp]
    \begin{center}
        \includegraphics[width=10cm]{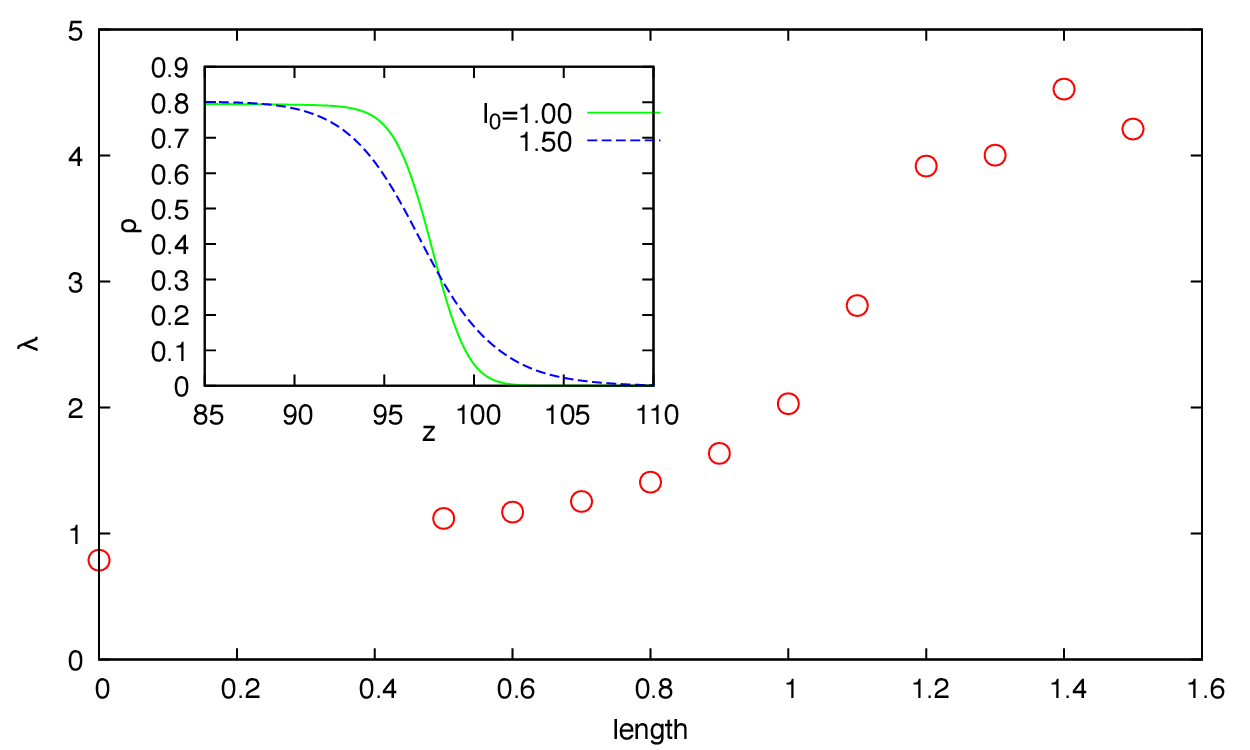}
        \caption{The bond length dependence on the characteristic length $\lambda$ of the liquid-liquid interface. The length $\lambda$ is $0.788(4)$ without surfactant. Then the length increases as the bond length of the surfactant increases.
            As the bond length $l_0$ increases, the interface length $\lambda$ also increases and saturates at $l_0 \sim 1.2$.
            (Inset) The density profile of the liquid of A-atoms near the interface. By assuming the form $\tanh((z_c-z)/\lambda)$, we determined the characteristic length of the interface $\lambda$.}
        \label{fig:length_lam}
    \end{center}
\end{figure}

As shown in Fig.~\ref{fig:sur_water_rho}, the structure of the liquid-liquid interface depends on the bond length of the surfactants. To investigate the relation between the interface structure and the bond length of the surfactants, we determined the characteristic length of the interface. It is known that when two phases coexist across an interface, the density profile of each phase can be described by a hyperbolic tangent function~\cite{Watanabe2012}. The form of the local density of A atoms $\rho_A(z)$ is expected to be
\begin{equation}
    \rho_A(z) = \rho_\mathrm{w} \tanh((z_c - z)/\lambda),
\end{equation}
where $\rho_\mathrm{w}$ is the bulk density of the solvent, $z_c$ is the position of the interface, and $\lambda$ is the characteristic length of the interface, respectively. Here, We assume that the density of A atoms dissolved in a liquid of B atoms is negligible. We observed local densities for various bond lengths of surfactants, from which we determined the characteristic length of the interface. The bond length dependence on the characteristic length $\lambda$ is shown in Fig.~\ref{fig:length_lam}. The characteristic length in the absence of surfactant is plotted as $l_0=0$. While the length $\lambda$ is $0.788(4)$ without surfactant, the length increases as the bond length of the surfactant increases and saturates at $l_0 \sim 1.2$. Since the characteristic length at the interface corresponds to the correlation length of the solvent, it is suggested that the bond length of the surfactant works most effectively in lowering the interfacial tension when it is approximately the same as the correlation length of the solvent. This is why there is an optimal bond length for lowering interfacial tension as shown in Fig.~\ref{fig:length0.015_0.025}.

\subsection{The spectra of fluctuations}

\begin{figure}[htbp]
    \centering
    \includegraphics[width=7cm]{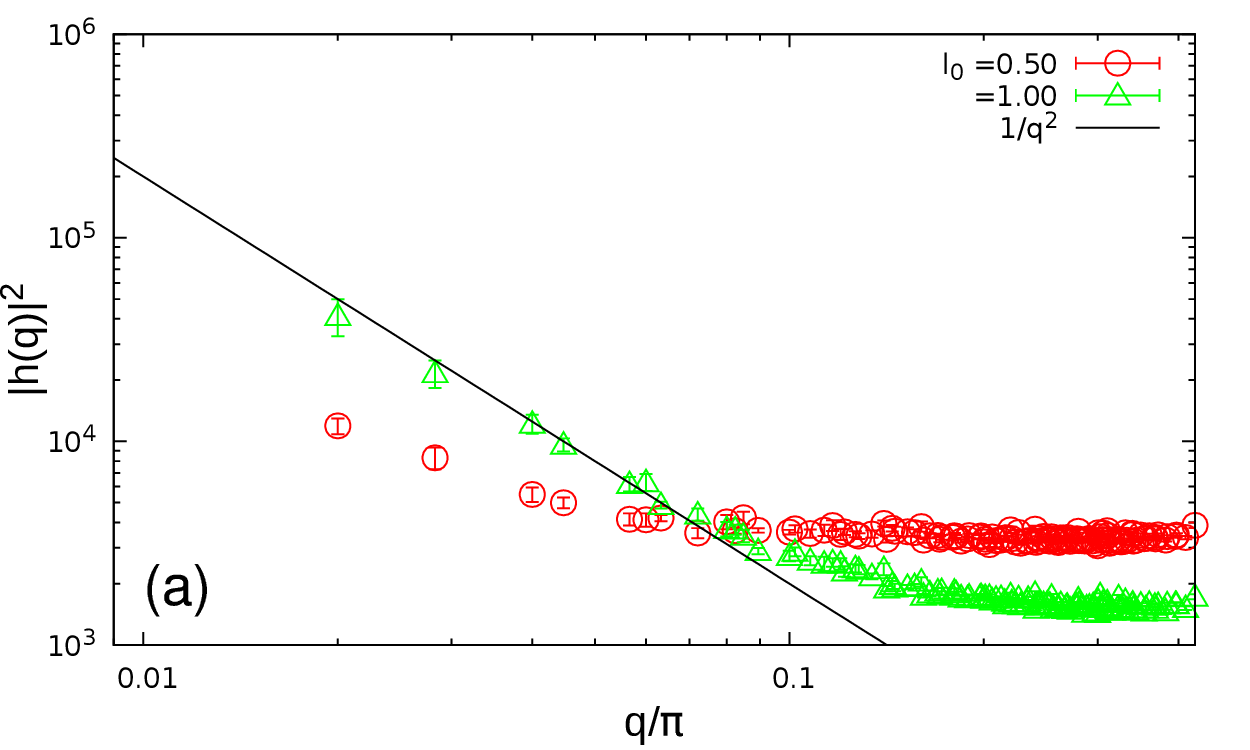}
    \includegraphics[width=7cm]{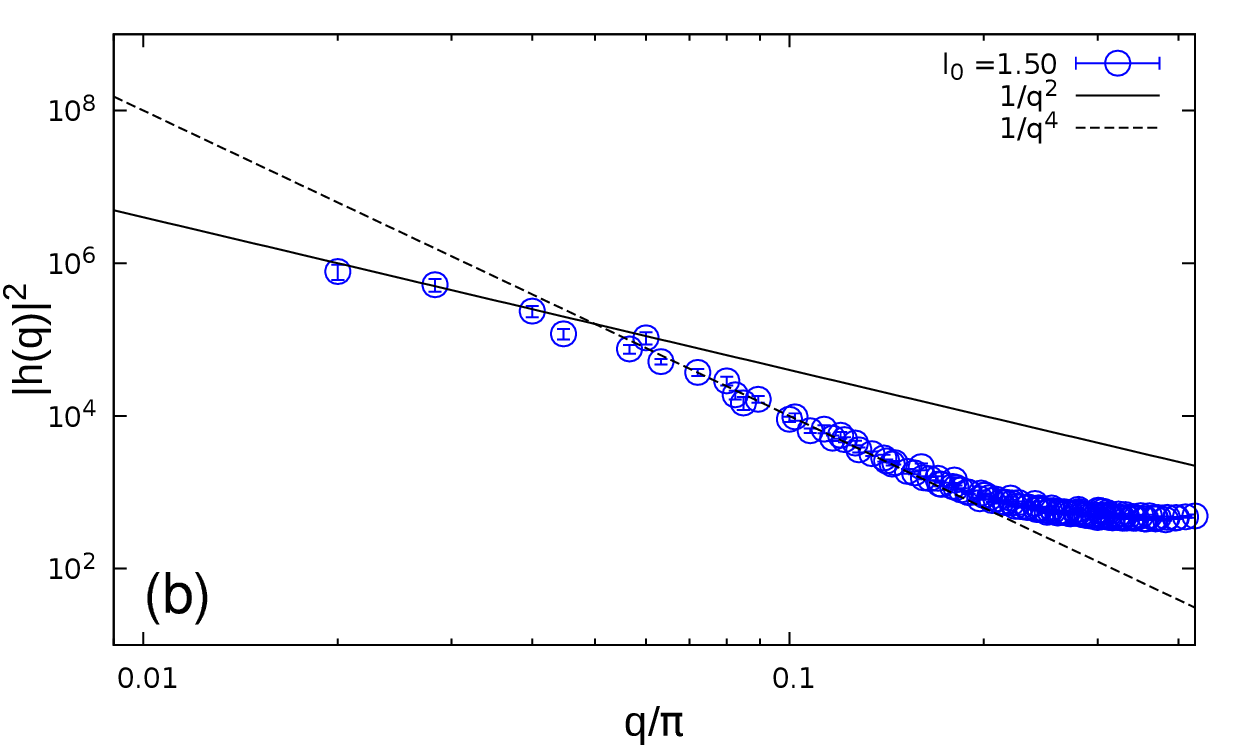}
    \caption{Spectra of the fluctuations of the surface structure for (a) the bond length $l_0 = 0.50$ and $1.00$ and (b) $l_0 = 1.50$. The solid lines denote $1/q^2$ and the dashed line denotes $1/q^4$ behavior. For $l_0 = 0.50$ and $1.00$, a behavior proportional to $q^2$ is observed for bond lengths of 0.50 and 1.00. For $l_0 = 1.50$, a crossover from $q^2$ to $q^4$ is observed.}
    \label{fig:fluctuation}
\end{figure}

\begin{figure}[htbp]
    \centering
    \includegraphics[width=7cm]{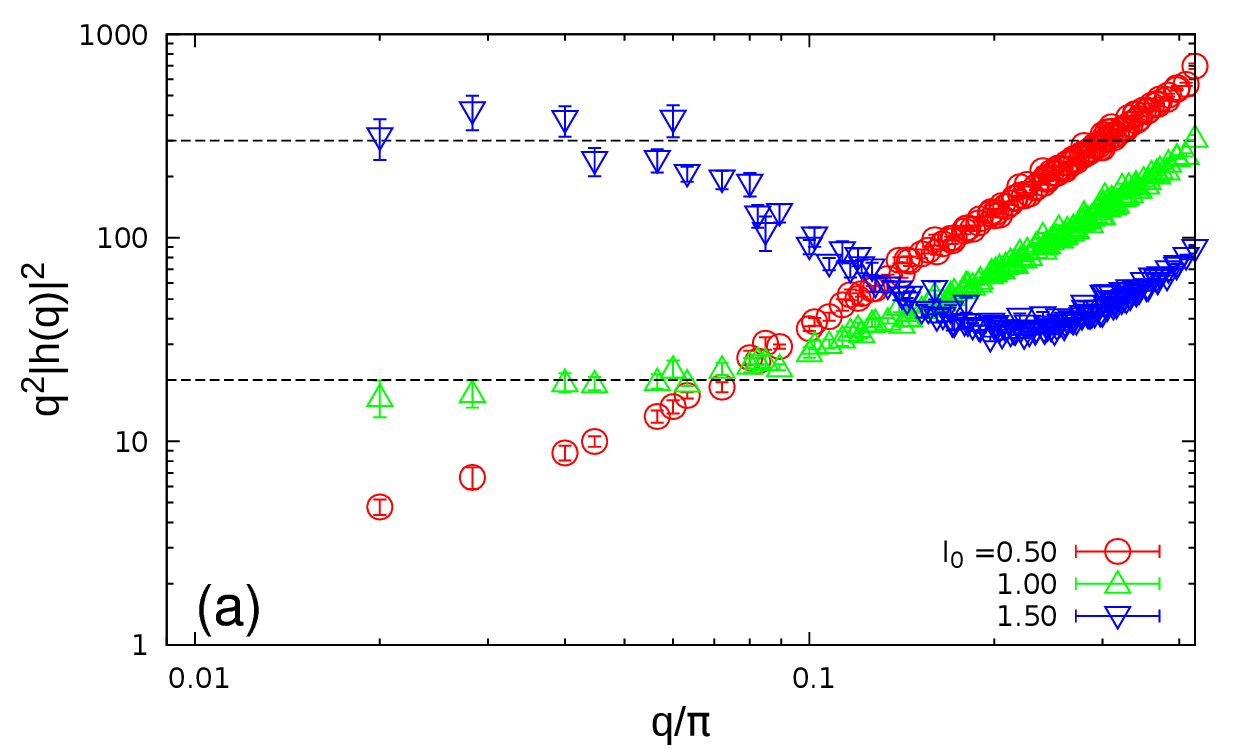}
    \includegraphics[width=7cm]{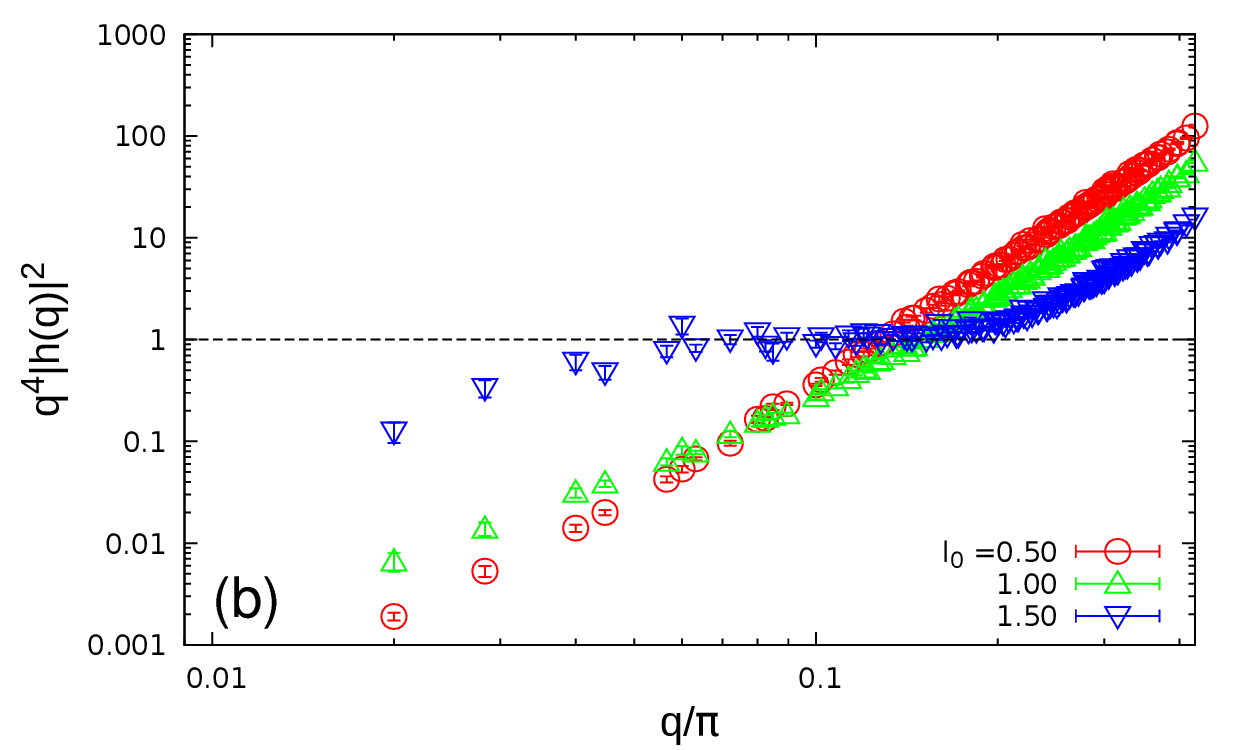}
    \caption{Graphs of (a) $q^2|h(q)|^2$ and (b) $q^4|h(q)|^2$. When the bond length $l_0 = 1.00$, $q^2$ is dominant because $q^2|h(q)|^2$ is constant in the region where $q$ is small. When the bond length $l_0 = 1.50$, $q^2|h(q)|^2$ is constant in the region where $q$ is small, and $q^4|h(q)|^2$ is constant when $q$ is about 0.1. This implies a crossover from $q^2$ to $q^4$.}
    \label{fig:q_fluctuation}
\end{figure}

For large bond lengths, the interfacial structure is maintained near the saturation concentration where the interfacial tension is virtually zero. Since the interfacial tension is zero, the bending rigidity is expected to be a restoring force for the membrane. Therefore, we investigate the spectra of fluctuation of the surface structure. Figure~\ref{fig:fluctuation} shows the Fourier spectra of the fluctuations. For small bond lengths ($l_0 = 0.50$ and $1.00$), the spectra show $q^2$ dependence in the low wavenumber region (Fig.~\ref{fig:fluctuation}(a)). In the high wavenumber region, the fluctuations are almost constant owing to thermal fluctuations. In contrast, when the bond length is large ($l_0 = 1.50$), a crossover from $q^2$ to $q^4$ is observed in the spectra as shown in (Fig.~\ref{fig:fluctuation}(b)). However, for ease of viewing, the graphs of $q^2|h(q)|^2$ and $q^4|h(q)|^2$ are shown in Fig.~\ref{fig:q_fluctuation}. From these graphs, it is clear that there is a region where $q^4$ dominate when the bond length is $1.50$. This is a crossover from the interfacial tension origin to the bending rigidity origin.
\section{Summary and Discussion}

We studied the effect of surfactants on the elasticity of the interface between two liquids. We modeled the surfactants as the simple diatomic molecules and we focused the bond length between the atoms of surfactants. Increasing the bond length would correspond to changing the size of the hydrophobic groups. However, the longer hydrophobic groups are more stabilized due to entropy effects. Additionally, longer hydrophobic groups yield a higher orientational order. By increasing the bond length of the two-bead model, only the positional relationship between hydrophobic and hydrophilic groups at the interface was changed without changing the orientational order.

We found that the interfacial tension depends on the surfactant's bond length. This finding suggests that the surfactant's structure can control the interfacial tension. Additionally, the interfacial structure was stable even when the interfacial tension was almost zero. We clarified that the restoring force of the interface originates from the bending rigidity when the interfacial tension is virtually zero. This result indicates that the restoring force of the interface switched from the interfacial tension to the bending rigidity. Since the surfactant molecules are sufficiently absorbed in the interface, the interfacial free energy is almost zero. Therefore, the bending rigidity is dominant in the restoring force of the interface. To the best of our knowledge, this is the first observation of the bending rigidity in a surfactant monolayer. The mechanical property of the system when surfactants are not saturated is one of the issues to be addressed in future studies.

These findings are useful for the design of Janus particles\cite{hu2012fabrication,ruhland2011janus}. If Janus particles that can change their length in response to external fields are designed, it may be possible to switch the interfacial tension. These could lead to medical applications, for example, a microcapsule that releases a drug from a capsule at any given time. The detailed design of such a device is the subject of future work.

\begin{acknowledgments}
    The authors would like to thank H. Noguchi and H. Nakano for fruitful discussions. This research was supported by JSPS KAKENHI, Grant No.~JP21K11923. The computation was partly carried out using the facilities of the Supercomputer Center, Institute for Solid State Physics (ISSP), University of Tokyo.
\end{acknowledgments}

\bibliography{reference}

\end{document}